\documentclass[11pt,preprint]{aastex}

\usepackage{graphicx}

\newcommand{\ha}{\hbox{H$\alpha$}}
\newcommand{\hb}{\hbox{H$\beta$}}
\newcommand{\hei}{\hbox{He\,{\sc i}}}
\newcommand{\hg}{\hbox{H$\gamma$}}
\newcommand{\hd}{\hbox{H$\delta$}}
\newcommand{\hi}{\hbox{H\,{\sc i}}}
\newcommand{\hii}{\hbox{H\,{\sc ii}}}
\newcommand{\oi}{\hbox{[O\,{\sc i}]}}
\newcommand{\oii}{\hbox{[O\,{\sc ii}]}}
\newcommand{\oiii}{\hbox{[O\,{\sc iii}]}}
\newcommand{\nii}{\hbox{[N\,{\sc ii}]}}
\newcommand{\sii}{\hbox{[S\,{\sc ii}]}}

\newcommand{\etal}{\hbox{et\thinspace al.\ }}

\righthead{Spectroscopic study of BCGs: II}
\lefthead{Kong et al.}

\begin{document}
\title{\LARGE \bf Spectroscopic Study of Blue Compact Galaxies\\
II. Spectral Analysis and Correlations}

\author{
X. Kong\altaffilmark{1,2},
F.Z. Cheng\altaffilmark{2},
A. Weiss\altaffilmark{1},
and
S. Charlot\altaffilmark{1,3}
}

\altaffiltext{1}{Max Planck Institute for Astrophysics, Karl-Schwarzschild-Str.
1, D-85741 Garching, Germany}
\altaffiltext{2}{Center for Astrophysics, University of Science and Technology
of China, 230026, Hefei, P. R. China}
\altaffiltext{3}{Institut d'Astrophysique de Paris, CNRS, 98 bis Boulevard
Arago, 75014 Paris, France}

\date{\centering Accepted for publication in A\&A}

\begin{abstract}
This is the second paper in a series studying the star formation 
rates, stellar components, metallicities, and star formation 
histories and evolution of a sample of blue compact galaxies.  
We analyzed spectral properties of 97 blue compact galaxies, 
obtained with the Beijing Astronomical Observatory (China) 2.16 m 
telescope, with spectral range 3580\AA\ -- 7400\AA.  We classify 
the spectra according to their emission lines: 13 of the total 97 
BCGs sample are non-emission line galaxies (non-ELGs); 10 have 
AGN-like emission (AGNs), and 74 of them are star-forming galaxies 
(SFGs). Emission line fluxes and equivalent widths, continuum 
fluxes, the 4000 \AA\ Balmer break index and equivalent widths of 
absorption lines are measured from the spectra.

We investigate the emission line trends in the integrated spectra 
of the star-forming galaxies in our sample, and find that: 
1) The equivalent widths of emission lines are correlated with the 
galaxy absolute blue magnitude $M_B$; lower luminosity systems tend 
to have larger equivalent widths.  
2) The equivalent width ratio \nii6583/\ha\ is anti-correlated with 
equivalent width \ha; a relationship is given that can be used to 
remove the \nii\ contribution from blended \ha+\nii6548, 6583.  
3) The \oii, \hb, \hg\ and \ha\ fluxes are correlated; those can be 
used as star formation tracers in the blue.  
4) The metallicity indices show trends with galaxy absolute magnitude 
and attenuation by dust; faint, low-mass BCGs have lower metallicity 
and color excess.
\end{abstract}


\section{Introduction}

Blue compact galaxies (BCGs) are characterized by their very blue color, 
compact appearance, high gas content, strong nebular emission lines, and 
low chemical abundances (Kunth \& {\" O}stlin 2000; \"{O}stlin et al. 2001). 
These properties are typical of unevolved systems, thus suggesting that 
BCGs should have suffered very few bursts of star formation during their 
lives and that some of them are probably experiencing their first burst.  
In a recent review, Kunth \& \"{O}stlin (2000) argued that, despite a few 
remaining young galaxy candidates (like I Zw 18, SBS 0335-052), in most 
BCGs an old underlying stellar population does exist, revealing at least 
another burst of star formation (SF) prior to the present one 
(Papaderos \etal 1996; Kong \& Cheng 2002). 
In addition, these properties make BCGs represent an extreme environment 
for star formation that differs from that in the Milky Way and in other 
quiescent nearby galaxies (Izotov \& Thuan 1999; Izotov et al. 2001).
Detailed studies of BCGs are important not only for understanding their 
intrinsic properties, but also for understanding of the chemical evolution 
of galaxies, for constraining models of stellar nucleosynthesis, for 
understanding star formation processes and galaxy evolution in different 
environments.  

To measure the current star formation rates, stellar components, 
metallicities, and star formation histories and evolution of BCGs, we 
have prepared an atlas of optical spectra of the central regions of 97 
blue compact galaxies in the first paper of this series (Kong \& Cheng 2002; 
Paper I). Because we want to combine the optical spectra we obtained 
with \hi\ data to constrain simultaneously the stellar and gas contents 
of BCGs, we selected most of our sample from \hi\ surveys by Gordon \& 
Gottesman (1981). The spectra were obtained at the 2.16 m telescope at 
the XingLong Station of the Beijing Astronomical Observatory (BAO) in 
China. A 300 line mm$^{-1}$ grating was used to achieve coverage in the 
wavelength region from 3580 to 7400 \AA\ with about 10\AA\ resolution.

In the present paper, we provide measurements of emission line equivalent
widths and fluxes, equivalent widths of absorption lines, 4000 \AA\, 
Balmer break index, as well as fluxes at several points of the continuum 
for our BCGs sample. We explore the trends in emission line fluxes and 
equivalent widths in the integrated spectra of SFGs in the sample. The 
absorption features and the continuum colors will be used to study the 
stellar population components and star formation history of blue compact 
galaxies.  
The emission line equivalent widths and fluxes will be used to determine 
the physical parameters of blue compact galaxies.

The paper is organized as follows. 
In Section 2, we classify spectra according to their emission lines.  
Emission line equivalent widths and fluxes measurements are presented in 
Section 3.  
Section 4 describes the continuum determination and the measurements of 
stellar absorption equivalent widths.  
In Section 5 we present an analysis of the emission line equivalent widths, 
line ratios and blue magnitudes of BCGs. 
Section 6 summarizes our conclusions.

\section {Spectral Classification}

Most of our BCGs sample was selected from the previous neutral 
hydrogen studies of blue compact galaxies by Gordon \& Gottesman 
(1981). This study has focused on the Haro, Markarian, and Zwicky 
lists of galaxies and hence objects were selected on the basis of 
a blue color, UV-excess or compactness, but not on the basis of 
emission line strength (Smoker et al. 2000).  The optical spectral 
observations of these galaxies show a range in spectral properties; 
from galaxies with absorption line spectra to narrow emission line 
objects classified as SFGs. 

In order to study our sample galaxies in detail, we use emission 
lines to classify the sample spectra into three types: 
non-emission line galaxy (non-ELG), low-luminosity active galactic 
nuclei (AGN) and star-forming galaxy (SFG). Our classification 
scheme is outlined below.

Because the \ha\ recombination line is easily detected in optical 
spectrum and only weakly affected by dust and underlying stellar 
absorption, we first separated the spectra into two broad categories, 
emission line galaxy and non-emission line galaxy, using the \ha\ 
recombination line.  When \ha\ is detected in emission, we classified 
the galaxy as an emission line galaxy, otherwise as a non emission 
line galaxy.  13 of 97 BCGs spectra have no \ha\ recombination 
emission lines, and were classified as non-ELGs.  Stellar \ha\ 
absorption lines are, in fact, prominent in all of these spectra.
 
Next we classify the remaining 84 emission line spectra into active 
galactic nuclei and star-forming galaxies based on the next 3 steps: 

\begin{enumerate} 
\item
Because of their very large permitted line widths, Seyfert 1 galaxies 
are easily recognizable.  We classify 5 galaxies (Mrk 335, Mrk 352, 
Mrk 6, I Zw 26, and Mrk 50) with broad emission lines in their spectra 
as Seyfert 1 galaxies.

\item 
To discriminate the Seyfert 2 from star-forming galaxies, we 
consider the intensity ratio (corrected for reddening, see Sec. 3.4) 
$R_{23} = (\oii3727 + \oiii4959,5007)/\hb$ (Dessauges - Zavadsky et al. 
2000) to 
the remaining 78 galaxies except for VII Zw 631.  Dessauges - 
Zavadsky et al. (2000) showed empirically that galaxies with log 
$R_{23} \simeq 1.1 $ are mostly Seyfert 2.  Based on this diagnostic, 
we classify 4 galaxies, Mrk 1, Mrk 622, Mrk 198, Mrk 270, as Seyfert 
2. The classification of Seyfert 1 and Seyfert 2 agrees with the 
results of V\'{e}ron-Cetty \& V\'{e}ron (2001).

\item 
When neither \oii3727 nor \oiii4959,5007 are available, we use the 
flux ratio of \ha\ and \nii6583 to identify an AGN (Carter et al. 2001).  
If $\log$($\nii6583/\ha$) $<$ $-0.25$, the galaxy is classified as 
a star-forming galaxy (SFG, \hii-like); otherwise it is classified 
as an AGN.  This allows us to classify VII Zw 631, which has 
$\log$($\nii6583/\ha$)= $-0.07$, as an AGN. 

\end{enumerate}

To summarize, our sample consists of 13 non-ELGs; 10 AGNs; and 74 
SFGs.

\section{Emission line measurements}

\begin{figure*}
\centering
\includegraphics[angle=-90,width=\textwidth]{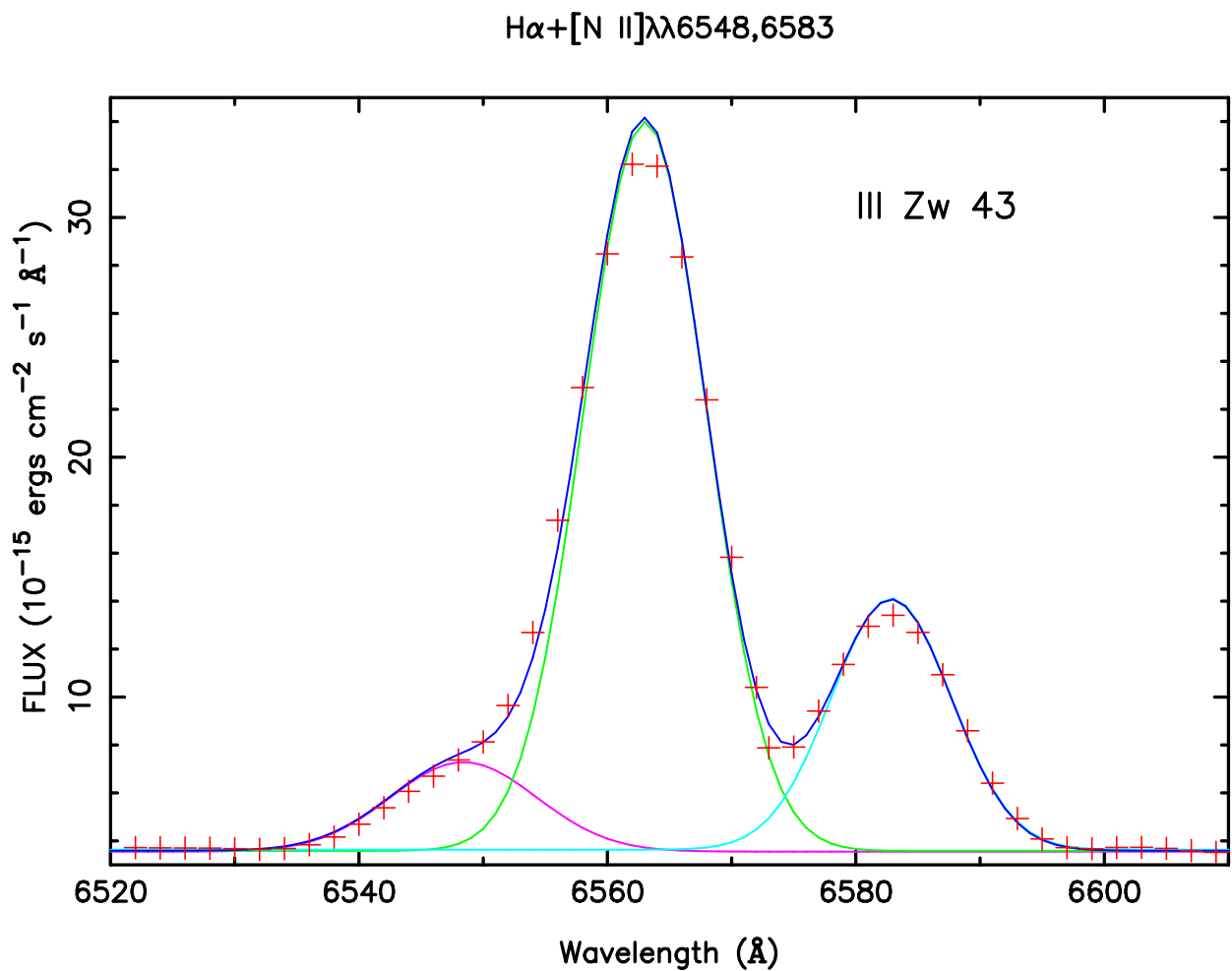}
\caption{
Example of Gaussian fit for the blended lines \ha+\nii6548, 6583. 
Three narrow Gaussian components are used. The crosses are observed 
spectral data.}
\label{fig1}
\end{figure*}

\subsection{Measurements}

The prominent spectral features of SFGs and AGNs include some 
commonly found strong emission lines, such as \oii3727, \hb4861, 
\oiii4959,5007, \ha6563, \nii6583, \sii6716, 6731, and some less 
common emission lines, such as \hg4340, \oiii4363, \hei5876, and 
\oi6300.

The rest-frame equivalent widths, $EWs$, and integrated fluxes, $F$, 
of the emission lines were measured by direct numerical integration, 
using the SPLOT program in IRAF. The continuum levels and 
integration limits for the lines were set interactively, with repeat 
measurements made in difficult case.  For single emission lines 
such as \hb4861, \oiii5007, direct integral methods were used. This 
method allows the measurement of lines with asymmetric shapes (i.e. 
with deviations from Gaussian profiles).  For blended lines such 
as \ha, \nii6548,6583, and the \sii6716, 6731 doublet, we used the 
Gaussian deblending program of SPLOT.  In Figure 1, as an example, 
we show the three narrow Gaussian components to fit of \ha, 
\nii6548,6583 of III Zw 43.  Note that in these blended cases, the 
lines are only partly blended. The interactive method allows us to 
control by eye the level of the continuum, taking into account 
defects that may be present around the line measured.  It does not 
have the objectivity of automatic measurements, but it does allow 
us to obtain reliable, accurate measurements.

The equivalent widths of various emission lines are listed 
in Table 1, for all SFGs and active galactic nuclei.  
The objects are ordered by increasing right ascension at the epoch 
2000 ($\alpha_{2000}$).  
Column (1) lists the galaxy name (same as Table 1 of Paper I).  
Columns (2) -- (9) list the equivalent widths of the commonly found 
emission lines. 
Columns (10) -- (13) list the equivalent widths of less commonly found 
emission lines.  
The second line for each entry lists an estimate of the error (see in 
Sec 3.2).  
We use the convention that positive equivalent widths denote emission 
to conserve space and improve readability. A dash in the table indicates 
either that the corresponding segment of the spectrum is lacking or that 
the spectrum was too noisy in the region to give a reliable value of 
equivalent widths. 
We have chosen an equivalent width of 1.0 \AA\ as the lower limit for 
true detection.  The observed emission line fluxes (the Galactic 
foreground reddening were corrected, see Kong \& Cheng 2002) are listed 
in Table 2.

\subsection{Standard deviations}

\begin{figure*}
\centering
\includegraphics[angle=-90,width=\textwidth]{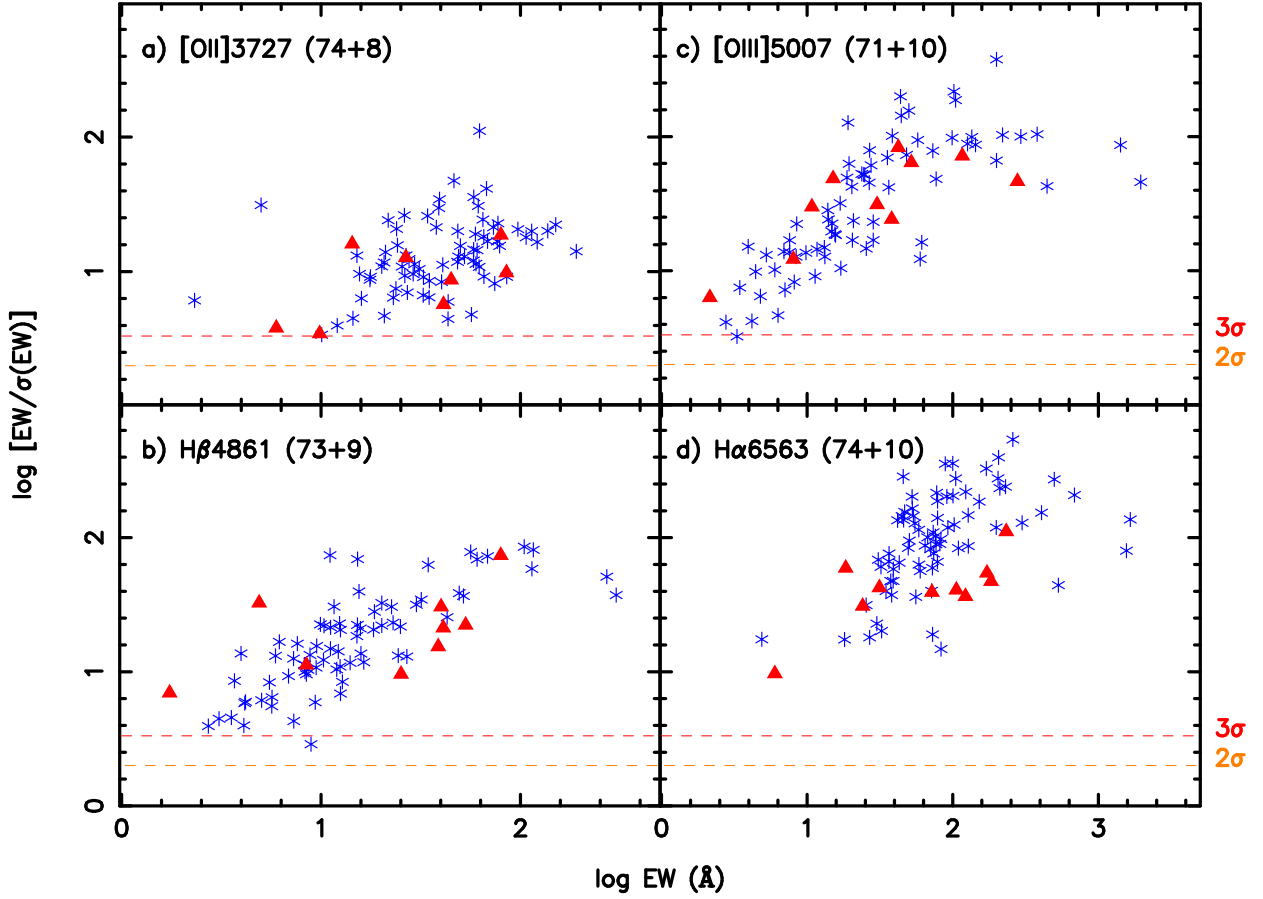}
\caption{
Log of the detection level of equivalent widths, $EW$ i.e. 
log~($EW$/$\sigma$($EW$)), versus log $EW$ for 4 emission lines. 
The emission line name, the number (SFG + AGN, $EW > 1.5$\AA) of plotted 
data and the 2$\sigma$, 3$\sigma$ detection levels (dashed lines) are 
indicated in each panel.  
The plotting symbols are coded according to spectral classification, 
the asterisks correspond to SFGs, the 
triangles to AGNs.}
\label{fig2}
\end{figure*}

For measurements of emission lines and absorption lines where the 
slope and curvature of the continuum are well defined, the main 
sources of random errors in the flux and equivalent width 
measurements are the uncertainty of the overall height of the 
continuum level, the individual intensity points within the 
interval of integration, the signal-to-noise ratio of the continuum, 
and the uncertainty in the choice of the best-fitting profile 
parameter.  To estimate $1\sigma$ standard deviations of emission 
lines, we followed the method outlined in Tresse et al. (1999), based 
on the formulae of propagation of errors and Poisson statistics.  
The derivation of error formulae can also be found in Longhetti et 
al. (1998).

The error $\sigma_{F}$ in the flux $F$ of an emission line can be 
expressed as (Tresse et al. 1999):

\begin{equation}
\sigma_{F}  =  \sigma_{c} D \sqrt{2 N_{pix} + EW / D } 
\end{equation}

The error $\sigma_{EW}$ with equivalent width $EW$ of the line 
can be expressed as:

\begin{equation}
\sigma_{EW} = \frac{EW}{F} \sigma_{c} D \sqrt{EW / D 
+ 2 N_{pix} + (EW / D)^2 / N_{pix} }
\end{equation}
where $\sigma_{c}$ is the mean standard deviation per pixel of the 
continuum on each side of the line, $D$ is the spectral dispersion 
in \AA\ per pixel (for our spectral sample, D=4.8) and $N_{pix}$ 
is the number of pixels covered by the line.  Because the 
signal-to-noise ratio for each pixel was estimated by scaling the 
continuum variance according to Poisson statistics, these are not 
exactly the formal statistical $1\sigma$ errors of $F$ or $EW$. Our 
approximation slightly overestimates the errors, but in our 
analysis, we are mainly interested in the consistency of the 
estimation of errors.

In Figure~\ref{fig2}, we plot a logarithm of the detection level 
of equivalent widths, log~($EW/\sigma(EW)$), versus log $EW$ for 
4 commonly found emission lines, \oii3727, \hb, \oiii5007, and \ha\ 
with equivalent widths above 1.5\AA.  
The result shows our equivalent widths limit is at a $3\sigma$ 
confidence level for those commonly found emission lines, and the 
typical uncertainty in these measurements is less than 10\%.  For 
those less common lines, such as \hg4340, \oi6300, the measurements 
typically have confidence levels $\geq 2\sigma$, and the typical 
uncertainty in these measurements is about 20\%.

\subsection{Stellar absorption correction}

Fluxes of emission lines will be used to determine the internal 
reddening of emission line regions, the star formation rate, and 
the element abundance of galaxies.  It is known that the 
measurements are an underestimate of the real flux of the spectral 
lines, because of the underlying absorption component.  To correct 
the underlying stellar absorption, some authors (such as Popescu \& 
Hopp 2000) adopt a constant equivalent width ($1.5-2$ \AA) 
for all the hydrogen absorption lines.  
Because the real value of the absorption equivalent width is
uncertain and dependent on the age of star formation burst and star
formation history (Izotov et al. 1994; Gonz{\' a}lez Delgado et al. 1999), 
the other usual correction 
for the contamination by stellar absorption lines assumes absorption 
equivalent widths, and iterates until the color excesses derived from 
\ha/\hb, \hb/\hg, and \hb/\hd\, ratios converge to the same value 
(Izotov et al. 1994).  

To derive the absorption equivalent width for hydrogen lines, we 
have applied an empirical population synthesis method, which uses 
observed properties of star clusters as a {\it base} 
(Cid Fernandes et al. 2001), 
to our BCG spectra.  This empirical population synthesis method can 
give the synthetic stellar population spectrum, so we can measure 
these underlying stellar absorption features for hydrogen lines. A 
full description of this application and equivalent widths of 
underlying stellar absorption lines will be presented in a 
forthcoming paper.

Emission line fluxes of \ha, \hb, and \hg\, are corrected for this 
underlying absorption effect as follows:

\begin{equation}
F^{\rm cor}_{\rm line}=F^{\rm obs}_{\rm line}(1 + 
EW^{\rm abs}_{\rm line}/EW^{\rm obs}_{\rm line}), 
\end{equation}

where $F^{\rm cor}_{\rm line}$ and $F^{\rm obs}_{\rm line}$ (see 
in Table 2, were corrected for Galactic extinction) are respectively 
the absorption corrected and the observed emission line fluxes, and 
$EW^{\rm abs}_{\rm line}$ and $EW^{\rm obs}_{\rm line}$ (see in Table 
1) are, respectively, the equivalent widths of the underlying 
stellar absorption line and of the observed emission line.

\subsection{Dust attenuation corrections}

The extinction of interstellar dust in SFGs 
modifies the spectra of these objects.  It is necessary to correct 
all observed line fluxes for this  internal reddening.  The most 
widely method used to correct the emission line spectra for the 
presence of dust is based on the relative strengths of low order 
Balmer lines. In order to have an internally 
consistent sample, we applied this method to each of our objects, 
using only the ratio of the two strongest Balmer lines, \ha$/$\hb.

We used the effective absorption curve $\tau_{\lambda} = 
\tau_V(\lambda/5500{\rm \AA})^{-0.7}$, which was introduced by 
Charlot \& Fall (2000). 
The color excess arising from attenuation by dust in a galaxy, 
$E^{int}_{B-V}$, can be written:

\begin{equation}
E^{int}_{B-V} = A_V/R_V = 1.086 \tau_V/R_V  ,
\end{equation}

\begin{equation}
\tau_V = 
-\frac{{\rm ln}[F(\ha)/F(\hb)]-{\rm ln}[I(\ha)/I(\hb)]}
{[(\lambda_{\rm H\alpha}/5500)^{-0.7} - (\lambda_{\rm 
H\beta}/5500)^{-0.7}]}
\end{equation}
where $I(\ha)/I(\hb)$ is the intrinsic Balmer flux ratio, 
$F(\ha)/F(\hb)$ is the observed Balmer flux ratio (were corrected 
for Galactic extinction and underlying stellar absorption), 
$\tau_V$ is the effective $V$-band optical depth.  
$\lambda_{\rm H\alpha} =6563{\rm \AA}$, 
$\lambda_{\rm H\beta}  =4861{\rm \AA}$, , and 
$R_V=3.1$.  We adopted $I(\ha)/I(\hb)$ = 2.87 for SFGs 
, and $I(\ha)/I(\hb)$ = 3.10 for the AGN-like objects 
(Dessauges - Zavadsky et al 2000).  For 4 galaxies, I Zw 18, Haro 43, 
II Zw 70, and, I Zw 123, their observed flux ratio, F(H$\alpha$)/F(H$\beta$), 
are less than the theoretical flux ratio 2.87, we set $E^{int}_{B-V}$ to be 
zero.  
The results of color excesses  $E^{int}_{B-V}$ are listed in the last column 
of Table 3.

The value of the color excess was then applied to the observed 
spectrum, and the final, intrinsic line fluxes relative to \hb\ for 
each galaxy can be expressed as:

\begin{equation}
\frac{I(\lambda)}{I(H\beta)} =\frac{F(\lambda)}{F(H\beta)}
\ {\rm e}^{\tau_V [(\lambda/5500)^{-0.7} - (\lambda_{\rm 
H\beta}/5500)^{-0.7}]},
\end{equation}
where $F(\lambda)$ and $I(\lambda)$ are the dust-obscured (observed) 
and intrinsic line fluxes, respectively. 
The attenuation-corrected line intensities relative to \hb\ are given
in Table 3 for each star 
forming galaxy and AGN. The objects are ordered by increasing right 
ascension at the epoch 2000 ($\alpha _{2000}$).  Column (1) lists 
the galaxy name, columns (2) -- (12) list the line intensities 
relative to \hb, column (13) list the intrinsic flux of \hb, which 
was corrected for both internal and Galactic extinction, and 
underlying stellar absorption. Column (14) lists the color excesses  
$E^{int}_{B-V}$  for each galaxy.

\subsection{Comparison with previous studies}

Seven galaxies in our BCG sample  --- III Zw 43 (0211+038), II Zw 40 
(0553+033), Mrk 5 (0635+756), I Zw 18 (0930+554), Haro 4 (1102+294), 
Haro 29 (1223+487), I Zw 123 (1535+554) --- have been observed 
previously by  Izotov, Thuan \& Lipovetsky (1997, ITL97), Izotov 
\& Thuan (1998, IT98), and Guseva, Izotov, \& Thuan (2000, GIT00), 
with the Ritchey-Chretien spectrograph at the Kitt Peak National 
Observatory (KPNO) 4 m telescope, and with the GoldCam spectrograph 
at the 2.1 m KPNO telescope.  These high signal-to-noise ratio 
spectrophotometric observations allow us to test the quality of 
our data.  We perform a detailed comparison of these previous studies 
in this subsection.

In III Zw 43, GIT00 did not detect \oiii4363 line, the \sii6731 
line intensity ratio is about 10\% higher, and the other emission 
line ratios are in good agreement with ours.  
For II Zw 40 in GIT00, Haro 29 in ITL97, our data are 
in fairly good agreement with these studies.  In Mrk 5, our \oi6300 
and \sii6717\, line intensities are about 25\% higher, and \oii\, 
is $\sim$ 13\% lower than that in IT98.  In I Zw 18, our \hg, 
\oiii4363, \sii\, line intensities are stronger, but \hei5876 is 
weaker.  In Haro 4, some less strong lines are not good agreement 
with IT98.  
Finally for I Zw 123, the agreement is not good as the other
galaxies, our $I(\lambda)/I(\hb)$ data have large differences with 
ITL97, but the $F(\lambda)/F(\hb)$ are in good agreement with ITL97.  

\begin{figure}
\centering
\includegraphics[angle=-90,width=\columnwidth]{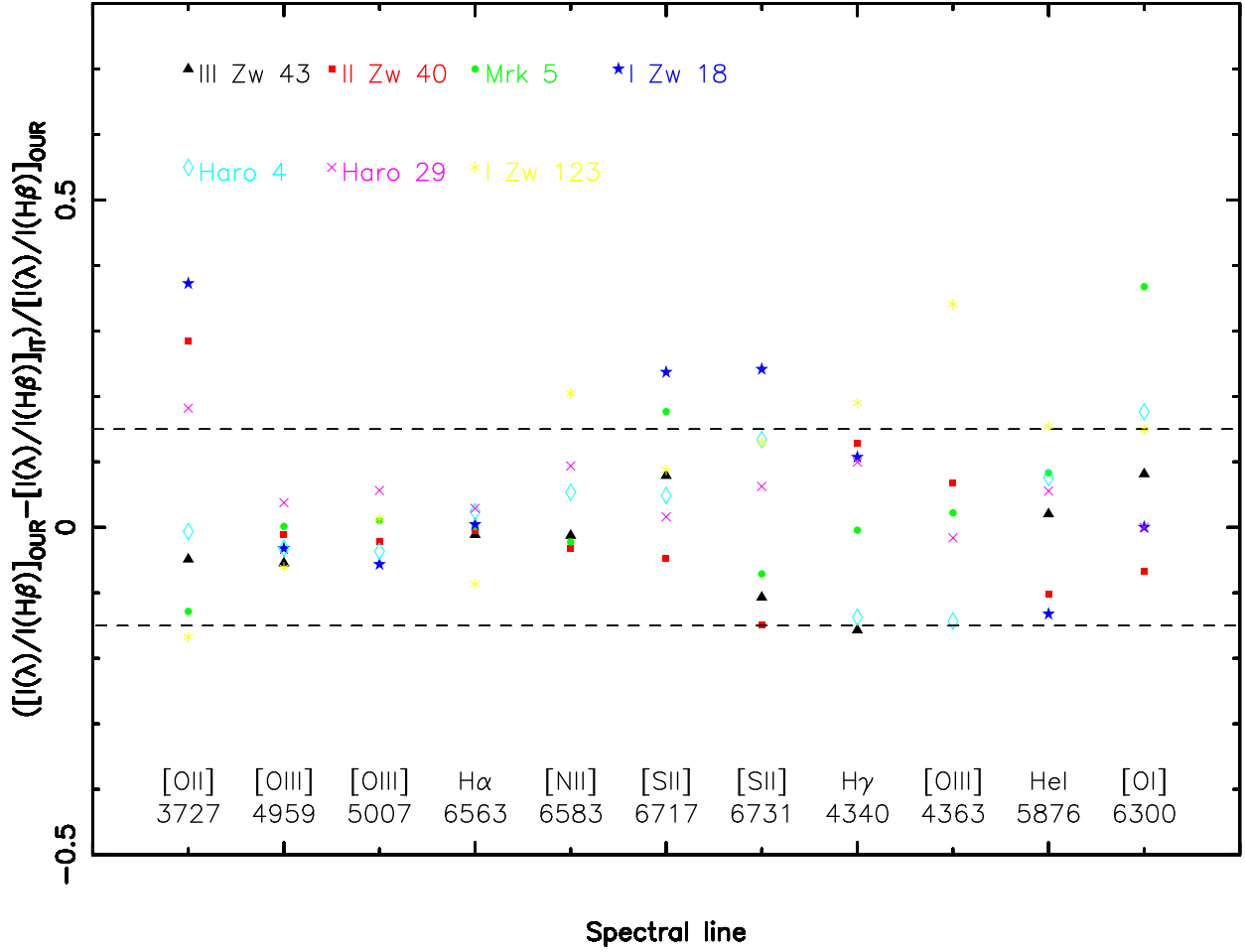}
\caption{
Difference between our line intensity ratios 
$[I(\lambda)/I(\hb)]_{\rm OUR}$ and these of previous studies 
$[I(\lambda)/I(\hb)]_{\rm IT}$.  The plotting symbols represent 
different galaxies, the dash lines outline the 15 per cent error 
window.}
\label{fig-com}
\end{figure}

We now display this comparison in a more visible form in Figure 
\ref{fig-com}.  The horizontal axis represent different spectral 
lines, the vertical axis shows the differences between our line 
intensities ($[I(\lambda)/I(\hb)]_{\rm OUR}$ ) and the values of GIT00, 
IT98 and ITL97 ($[I(\lambda)/I(\hb)]_{\rm IT}$).  We found, our line 
intensity ratios are in good agreement with these previous studies for 
most spectral lines of most galaxies, the difference between our 
sample and these studies is less than 10\% for those strong emission 
lines, and less than 15\% for those less strong lines, such as 
\hg4340,\oiii4363,\oi6300 of most galaxies.

The observed fluxes of \hb\ in our data are larger than those in 
previous studies, the explanation could be: 1) the data in these 
previous studies were not corrected for the Galactic extinction; 2) 
Our slit width is larger than that of previous studies; 3) the position 
angle of slit is different between ours and those previous studies. 
We will discuss the slit effect and derive an aperture correction 
for each galaxy in a future paper.  

\section{Continuum and absorption line measurements}

\begin{figure}
\centering
\includegraphics[angle=-90,width=\columnwidth]{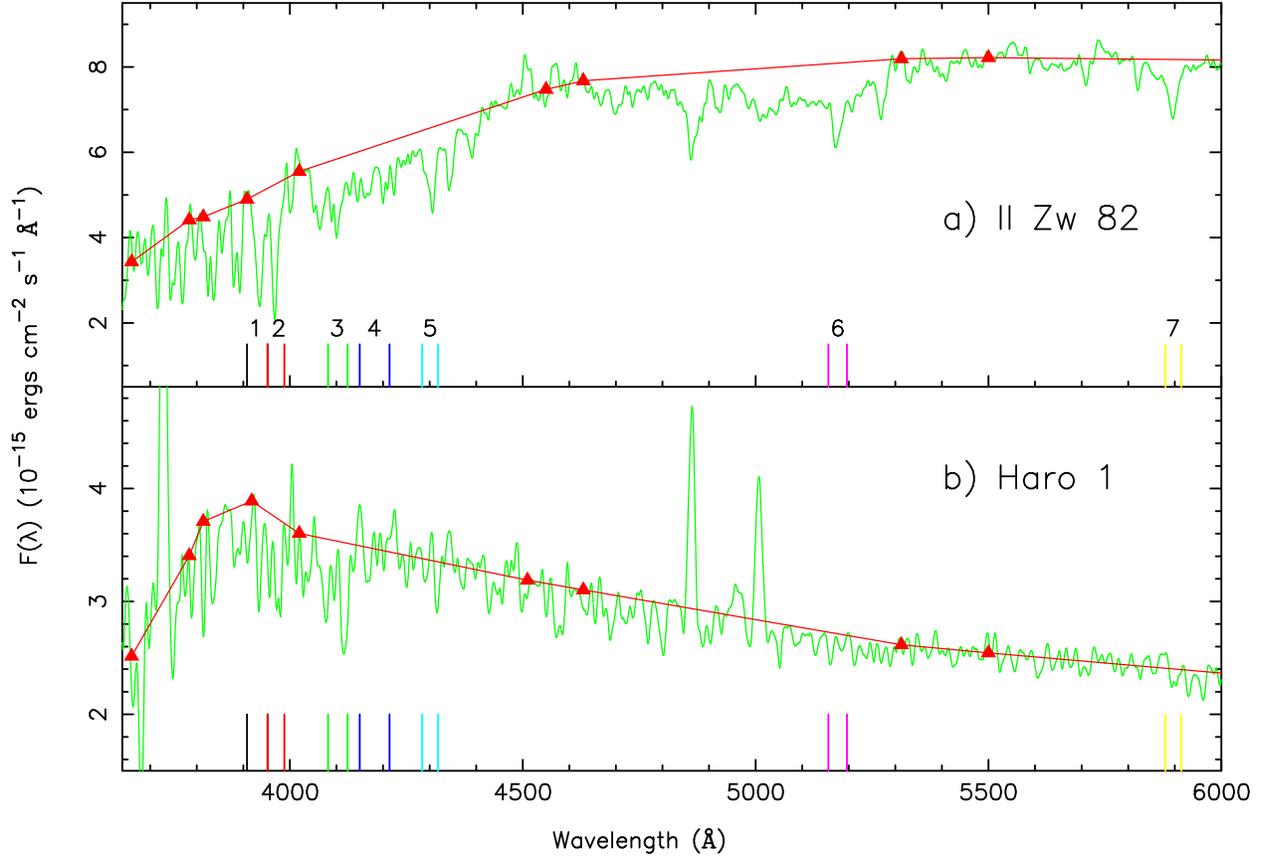}
\caption{
Illustration of the continuum determination procedure for the 
nuclear spectra of a) the non-emission line galaxy II Zw 82, and 
b) the star-forming galaxy Haro 1. The continuum pivot points are 
marked by the filled triangles. Vertical lines at the bottom of each 
panels indicate the wavelength windows used to measure the 
equivalent widths of Ca~K and H, H$\delta$, CN, G band, Mg and Na 
lines (See in Table 5).}
\label{fig4}
\end{figure}

The equivalent widths of absorption lines and the continuum colors 
provide information about the stellar populations and physical 
parameters of galaxies. One of our goals is to study the star 
formation history and chemical evolution of BCGs.  To this end, we 
also determined a pseudocontinuum at selected pivot points and 
measured the equivalent widths of 7 absorption features, 
integrating the flux within each window between the pseudocontinuum 
and the spectrum.  The absorption 
features and continuum points are chosen based on the  population 
synthesis method (Cid Fernandes et al. 2001) that will be used in a 
forthcoming paper.

\subsection{Continuum measurements}

In order to derive a continuum, we have measured the flux values 
at 9 pivot points, 3660, 4020, 4510, 4630, 5313, 5500, 6080, 6630, 
7043 \AA\, which were chosen to avoid regions of strong emission 
or absorption features (Bica 1988, Kong \& Cheng 1999, Saraiva et 
al. 2001). The corresponding fluxes were determined as averages in 
20\AA\ bins centered in the listed wavelengths.  The determination 
of the continuum has been checked interactively, taking into account 
the flux level, the noise and minor wavelength calibration 
uncertainties as well as anomalies due to the presence of emission 
lines.  The excellent quality of the spectra allowed a precise 
determination of the continuum in the majority of cases.

In addition, 3 point flux values (3784, 3814, 3918 \AA) were measured, 
which were necessary for the determination of the continuum in 
galaxies with strong contribution of late B to F stars which present 
several high-order Balmer absorption lines in this region (Bica et 
al. 1994).  Due to the crowding of the absorption lines, it is 
difficult to make automatic measurements. We thus selected the 
highest value of $\lambda$3784, $\lambda$3814 and $\lambda$3918 
\AA\ fluxes to represent the continuum in these spectral regions.

Figure~\ref{fig4} illustrates the application of the method to two 
of the sample spectra. The spectrum in Fig.~\ref{fig4}a corresponds 
to the nucleus of the non-emission line galaxy II Zw 82. The spectrum 
is purely stellar, so it is straightforward to measure the pseudocontinuum.  
Fig.~\ref{fig4}b shows the nuclear spectrum of Haro 1, 
a star-forming galaxy. The continuum is also well defined.  Overall, 
we find that the method works well for most spectra, with the 
exception of the nuclear regions of AGNs (see end of this Section).

The specific continuum wavelengths and corresponding fluxes for the 
non-ELGs and SFGs are shown in Table 4. The second 
line for each galaxy entry lists the errors with continuum 
measurements, calculated as the rms deviation from the average 
continuum flux.

\subsection{Absorption line measurements}

\begin{figure*}
\centering
\includegraphics[angle=-90,width=\textwidth]{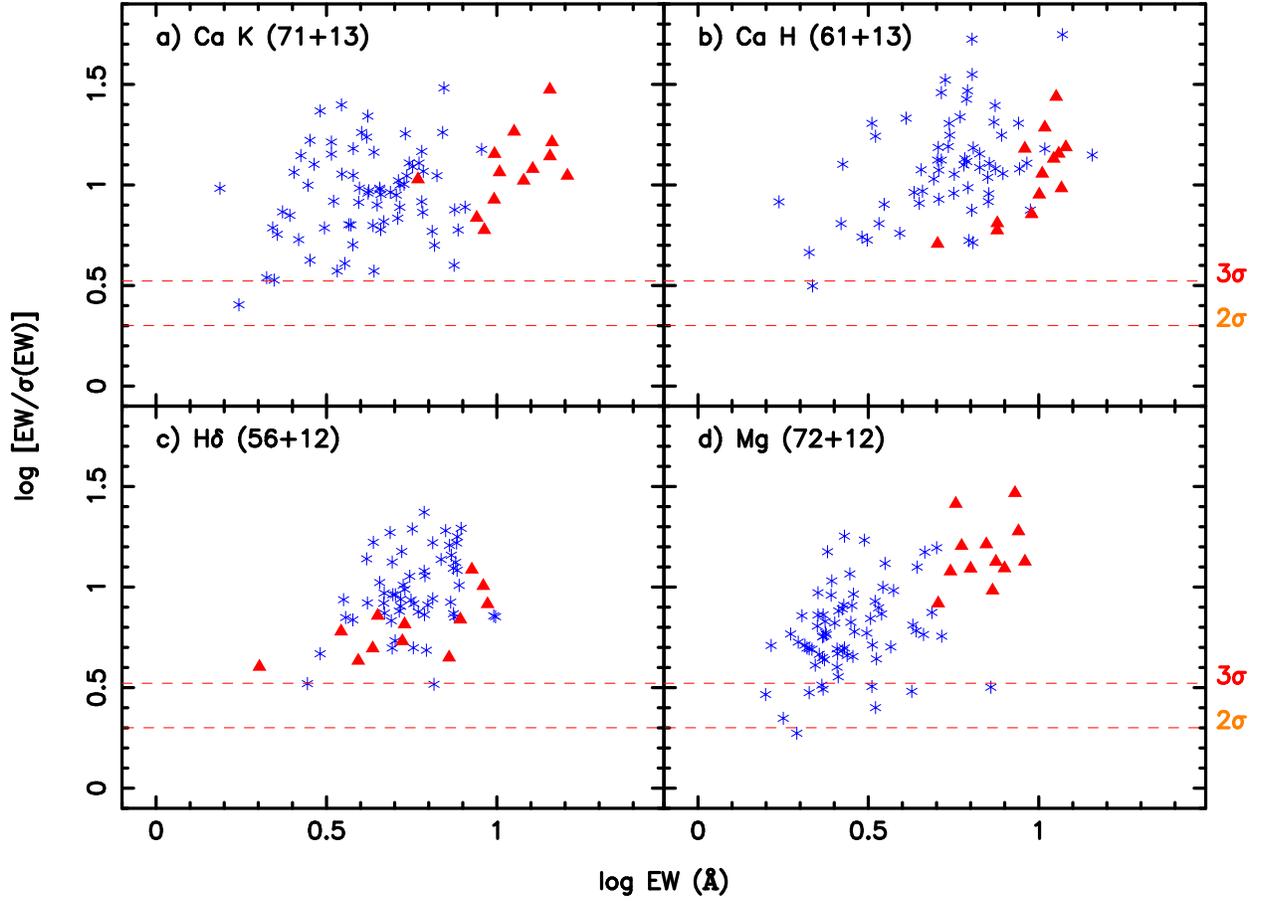}
\caption{
Log of the detection level of equivalent widths, i.e. log~[$EW/\sigma(EW)$], 
versus log $EW$ for 4 absorption lines. The absorption line name, the number 
(SFG + non-ELG, $EW > 1.5$\AA) of plotted data and the 2$\sigma$, 3$\sigma$ 
detection levels (dashed lines) are indicated in each panels. The plotting 
symbols are coded according to spectral classification, the asterisks 
correspond to star-forming galaxies (SFGs), the triangles to non-ELGs.}
\label{fig5}
\end{figure*}

\setcounter{table}{4}
\begin{table}
\begin{center}
\caption{Wavelength windows used to measure the equivalent widths of the
absorption lines.}
\centering
\begin{tabular}{ccll}
\hline
\hline
No. &Window   &Main                 &\\
    & (\AA)     &Absorber             &Identification\\
\hline
1&3908--3952& CaII K              &Ca K\\
2&3952--3988& CaII H$+$H$\epsilon$& Ca H\\
3&4082--4124& H$\delta$           & H$\delta$\\
4&4150--4214& CN                  & CN\\
5&4284--4318& $G$\ band           & $G$\ band\\
6&5156--5196& MgI$+$MgH           & Mg\\
7&5880--5914& NaI                 & NaI\\
\hline  
\end{tabular}
\end{center}
\end{table}

The spectrum of a galaxy is produced by the sum of the spectral 
characteristics of its stellar content (Weiss et al. 1995).  Observations 
of the integrated spectra of galaxies can be used to determine the 
distribution in age and metal abundance of the stellar population in these 
systems and hence to determine their epoch of formation and subsequent star 
formation history (Arimoto \& Yoshii 1986). To this aim, some strong, easily 
identifiable absorption features in our observed spectral range are measured, 
which include some age and metallicity sensitive absorption features. The 
absorption line names and adopted spectral wavelength windows are shown in 
Table 5.

The rest-frame equivalent widths of these absorption features 
were automatically computed by summing the observed fluxes below the 
continuum level, which itself is estimated by fitting a straight 
line to the fluxes in the above continuum regions.  The equivalent 
widths of the absorption features for these non-ELGs and 
SFGs are presented in Table 6.  
The second line for each galaxy entry lists the errors on equivalent 
widths, which are computed from equation (2) in this paper. 
Figure~\ref{fig5} shows that logarithm of the detection level of 
$EW$ versus log $EW$ for 4 absorption lines, such as Ca K, Ca H, \hd\, 
and MgI$+$MgH when its $EW > 1.5$\AA.  Our absorption feature $EW$ 
limit is at a $3\sigma$ confidence level for the absorption lines 
of most galaxies.  The mean uncertainty in these measurements is 
about 10\% for those non-ELGs and about 15\% for those SFGs.

\subsection{4000 \AA\ Balmer break index measurements}

As well as the absorption features, we also measured the 4000 \AA\ 
Balmer break. It is the strongest discontinuity in the optical 
spectrum of a galaxy and arises because of the accumulation of a 
large number of spectral lines in a narrow wavelength region 
(Bruzual 1983).  The main contribution to the opacity comes 
from ionized metals. In hot stars, the elements are multiple ionized 
and the opacity decreases, so the 4000 \AA\ break will be small for 
young stellar populations and large for old, metal-rich galaxies 
(Kauffmann et al. 2002).

We use the definition using narrower continuum bands than Bruzual's, 
which was introduced by Balogh et al (1999). The principal advantage 
of the narrow definition is that the index is considerably less 
sensitive to reddening effects.  It is defined as the ratio of the 
average fluxes (for frequency unit) measured in the spectral ranges 
4000--4100\AA\ and 3850--3950\AA: 
D4000vn=F$_{\nu}$[4000--4100\AA]$/$F$_{\nu}$[3850--3950\AA].  
The D4000vn index is simply a flux ratio and, hence the error is 
determined from standard propagation techniques. The D4000vn index 
value and its error for these non-ELGs and SFGs 
are presented in the last column of Table 6.

While for non-ELGs and SFGs the placement of the continuum is 
straightforward, this is not the case in the nuclear regions of most 
AGNs, where the numerous broad lines and intense non-stellar continuum 
complicate the analysis. The continuum points are impossible to 
determine accurately, and the equivalent widths of the absorption lines 
cannot be measured accurately. Therefore, for 10 AGNs, we only measured 
the integrated fluxes, $F$, and rest-frame equivalent widths, $EWs$, of 
the emission lines. We did not measure the continuum fluxes and 
equivalent widths of the absorption lines for those AGNs.  

\section{Emission line analysis for SFGs}

The main goal of this series of papers is to measure the current star 
formation rates, stellar components, metallicities, and star formation 
histories and evolution of BCGs.  For this purpose, we are mostly 
interested in the SFGs. 
Therefore, in present section, we do not consider the galaxies with 
Seyfert nuclei and the non-emission line galaxies in our BCG sample. In 
the following, the sample will refer to the 74 SFGs. We investigate the 
emission line trends in spectra of those SFGs in this section.

\subsection{Equivalent widths of emission lines}

\subsubsection{$EWs$ versus $M_B$}

It is interesting to explore how the equivalent widths of emission lines 
depend on the galaxy absolute blue magnitude $M_B$ (Column 6 of Table 1 
in Paper I).  The equivalent widths of \oii3727, \hb, and \ha\ are well 
correlated with $M_B$, the other emission lines are also correlated with 
$M_B$, but the spread in equivalent widths at a given luminosity is large. 
Lower luminosity systems tend to have larger equivalent widths for most 
of emission lines, except for \nii6583. 

\begin{figure}
\includegraphics[angle=-90,width=\columnwidth]{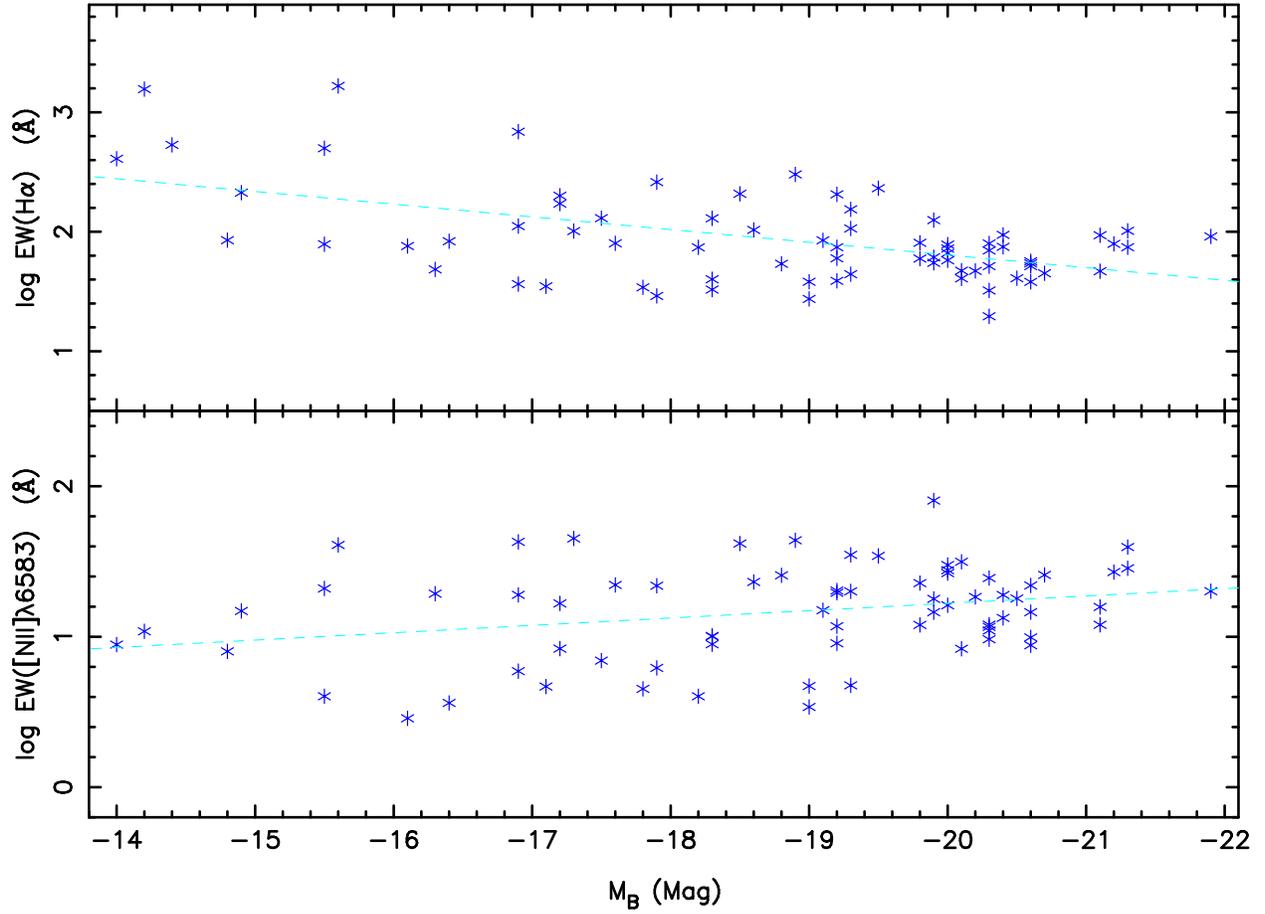}
\caption{
The logarithm of integrated \ha\ and \nii6583\, emission line $EWs$ 
plotted versus absolute $B$ filter magnitude.  A tendency towards 
larger \ha\ $EWs$ and smaller \nii\ $EWs$ in lower luminosity systems 
is seen.}
\label{eqw-amb}
\end{figure}

In the top panel of Figure \ref{eqw-amb}, we plot the \ha\ emission 
equivalent width as a function of $M_B$ for the SFGs in our sample.  
A pronounced trend towards larger equivalent widths at lower luminosities 
can be found, galaxies with the strongest \ha\ lines are of low luminosity. 
$EW$(\ha) is the ratio of the flux originating from UV photoionization 
photons ($ < 912$ \AA) to the flux from the old stellar population emitted 
in the rest-frame $R$ passband, which forms the continuum at \ha. Thus, a 
large equivalent widths is due either to a large UV flux (or $B$ absolute 
magnitude since they are correlated), or to a low continuum from old stars. 
In either case, this implies a blue continuum color. Hence, the observed 
trend of larger $EW$(\ha) for fainter galaxies implies that the faint SFG 
population is dominated by blue galaxies, while the bright SFG population 
is dominated by redder galaxies. 

In the bottom panel of Figure \ref{eqw-amb}, we plot the \nii6583\, 
emission line equivalent width as a function of $M_B$ for the SFGs 
in our sample. Its equivalent width behaves in the opposite way, 
lower luminosity systems tend to have smaller equivalent widths. 
Such a trend has also been found in other studies of nearby galaxies 
(Jansen et al. 2000). The global behavior of \nii6583\, $EW$ reflects 
intrinsic differences in the nitrogen abundance in BCGs, on average 
luminous BCGs are likely to be enhanced in nitrogen abundance.  This 
suggests that, in faint, low-mass, BCGs, nitrogen is a primary 
element, whereas in brighter, more massive BCGs it comes from a 
secondary source.

\subsubsection{$EW$(\nii)/$EW$(\ha) versus $EW$(\ha + \nii)}

\begin{figure}
\includegraphics[angle=-90,width=\columnwidth]{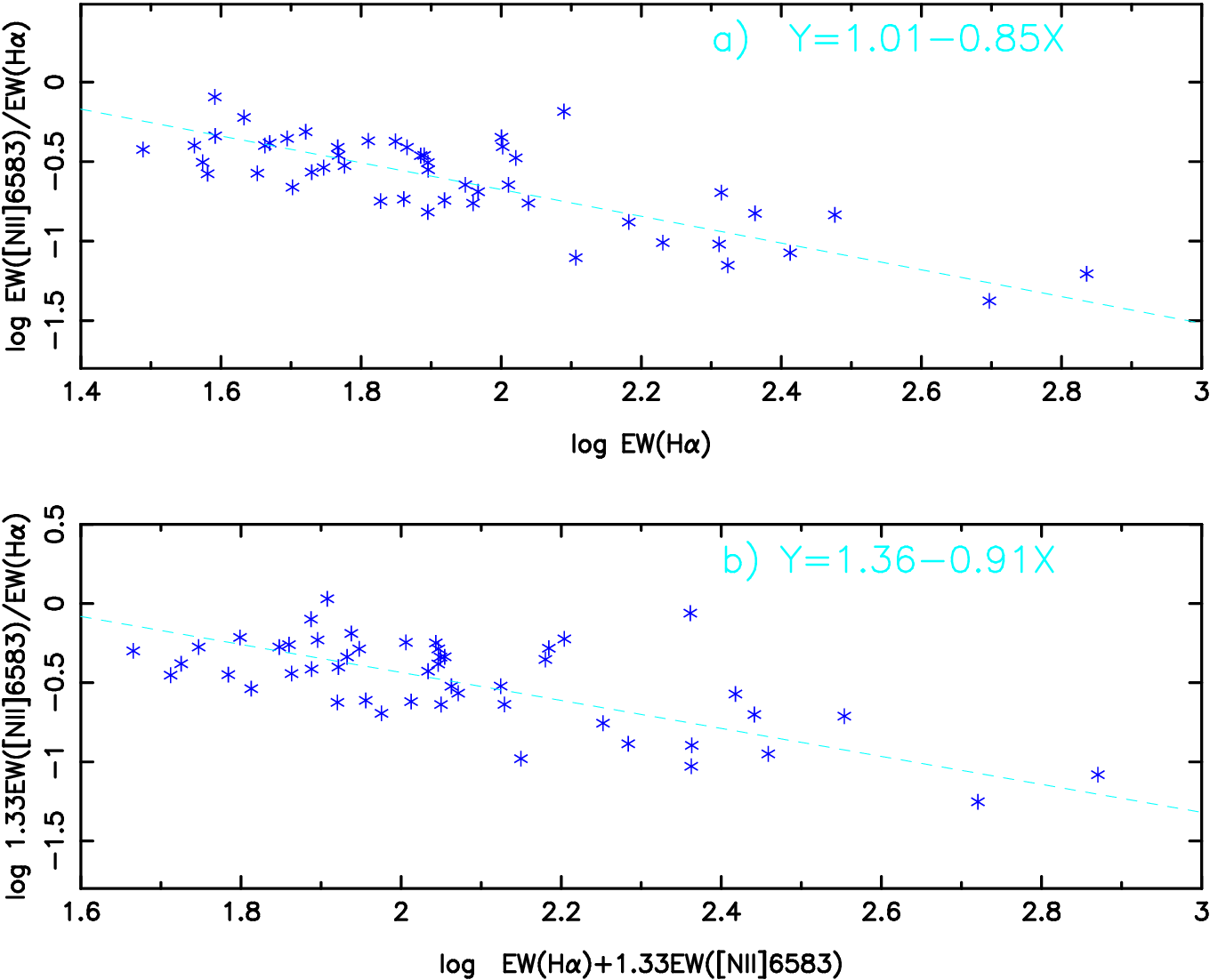}
\caption{
a) log $EW$(\nii6583)$ / EW$(\ha) versus log $EW$(\ha). 
b) log 1.33 $EW(\nii6583) / EW(\ha)$ versus log $EW(\ha) + 
1.33 EW(\nii6583)$.
}
\label{niiha}
\end{figure}

In deep large optical surveys, low-resolution spectroscopy or 
narrowband \ha\ imaging is often used. \ha\ and \nii6548, 6583 lines 
are often blended, so it is important to recover the flux solely 
in \ha\ to measure for instance the \ha\ luminosity function, hence 
to derive a star formation rate. Figure~\ref{niiha}a shows that the 
\nii6583/\ha\ $EW$ ratio decreases as a function of $EW$(\ha). All the 
spectra in this figure have  $EW$(\nii) and $EW$(\ha) $>$ 10 \AA, which 
can be measured very accurately. The \nii6583/\ha\ equivalent 
widths ratio is strongly correlated with $EW$(\ha).  A least-squares 
fit of this relation yields: log $EW$(\nii6583)$/EW$(\ha) = $(1.01 \pm 
0.14) - (0.85 \pm 0.07)$ log $EW$(\ha).

Since $EW(\nii6583) = 3EW(\nii6548)$, we also plot the relation 
1.33 $EW(\nii6583)$/ $EW(\ha)$ versus $EW(\ha) + 1.33 EW(\nii6583)$ in 
Figure~\ref{niiha}b. The trend is similar to that in 
Fig.~\ref{niiha}a, log $1.33 EW(\nii6583) / EW(\ha)$ = $(1.36 \pm 0.20) - 
(0.91 \pm 0.09)$ log $(EW(\ha) + 1.33EW(\nii6583))$.  Thus we can predict 
which value is expected for the ratio when observing the blend \ha\ 
+ \nii6548, 6583.  For instance, if this latter, EW(\ha) + 1.33EW(\nii6583), 
is $\sim100$ \AA, 
the ratio 1.33 $EW(\nii6583)$ / $EW(\ha)$ should be $\sim 0.35$.  The value 
of the ratio \nii6548,6583 / H$\alpha$, as determined by Kennicutt 
(1992), is usually taken to be 0.5 to remove the contribution of 
\nii\ to (\ha + \nii) blended emission.  This is slightly larger than 
the typical value for our star-forming galaxy sample trend, 
presumably because Kennicutt's sample contains a large fraction of 
early-type galaxies, which have systematically higher ratios 
(Tresse et al. 1999). 

\subsection{Fluxes of emission lines}

\begin{figure}[b]
\includegraphics[angle=-90,width=\columnwidth]{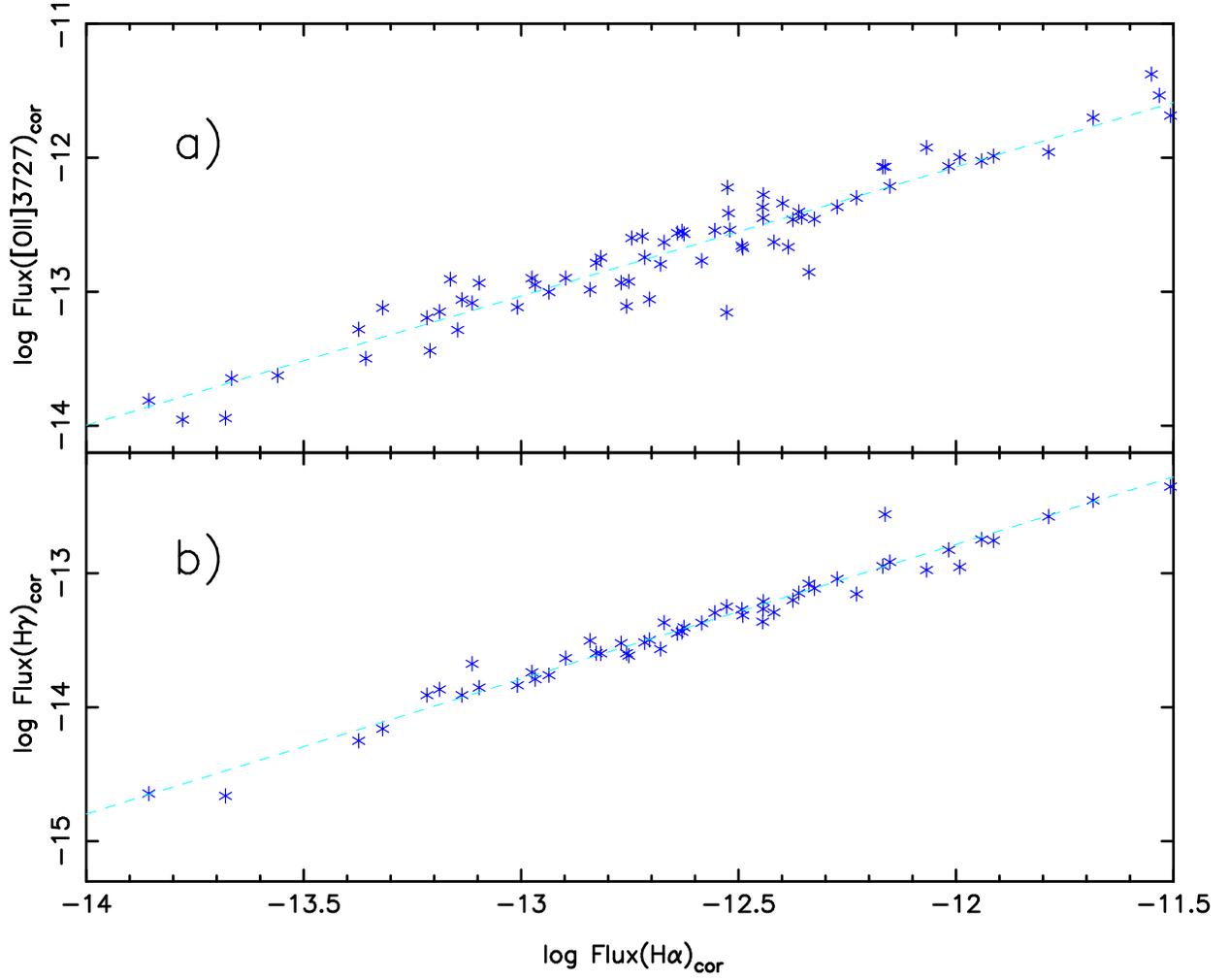}
\caption{
The correlation of the intrinsic (reddening and underlying absorption were 
corrected) emission line fluxes as a function of the intrinsic \ha\ 
emission line flux.  The dotted line indicates a {\bf linear} least-squares 
fit to the data points. Fluxes are in erg\, cm$^{-2}$\, s$^{-1}$.
a) \oii3727; b) \hg4340.}   
\label{flux-ha}
\end{figure}
  
Emission line fluxes are primary traces of the star formation rate 
(SFR) in galaxies (Kennicutt 1998). For star-forming galaxies, the 
Balmer emission line luminosities scale directly with the ionizing 
fluxes of the embedded young stars, and this makes it possible to 
use the Balmer lines to derive SFRs in galaxies. \ha\ is the 
best line for such applications, but beyond $z \simeq 0.2 - 0.3$, 
this line is redshifted into the near infrared.  To find other 
tracers of SFR we analyze the fluxes of other emission lines as a 
function of the intrinsic \ha\ flux.

Since the various prominent emission lines correlate with each other, 
any of them is likely to be a first order ranking indicator of SFR 
of star-forming galaxies, but the strongest correlations are found 
between \oii3727, \hg4340, \hb\ and \ha.  \oii3727 is the most 
useful star formation tracer in the blue.  In Figure \ref{flux-ha}a, 
we show the flux of \oii3727 as a function of the intrinsic \ha\ 
flux.  We found, as expected, these two lines have a strong 
correlation. 

From purely astrophysical considerations, the most reliable star 
formation tracers in the blue should be the higher order Balmer lines, 
since the fluxes of these lines scale directly with the massive star 
formation and are nearly independent of the temperature and 
ionization level of the emitting gas. Figure 
\ref{flux-ha}b shows the relation between the fluxes of \ha\ and 
\hg. A strong, roughly linear correlation between \ha\ and \hg\ is 
apparent.  
This correlation confirms that the \hg\ line can serve as a reliable 
star formation tracer in strong emission line galaxies, such as the 
SFGs in our BCG sample. In addition, the correlation between the 
intrinsic fluxes of \ha\ and \hb\ is stronger than that between the 
fluxes of \ha\ and \hg, \hb\ is another good star formation tracer 
for SFGs.

Recently, Charlot \& Longhetti (2001) quantified the uncertainties 
in these SFR estimators. They found SFR estimates based purely on 
one of emission line luminosity of galaxies could be in error by 
more than an order of magnitude. On the other hand, with the help 
of other emission lines, these errors can be substantially reduced.  
Based on our high quality spectrophotometric data, we can derive 
star formation rate for each galaxy, using these different star 
formation rate estimators.

\subsection{Line ratios and metallicity}

\begin{figure*}
\centering
\includegraphics[angle=-90,width=\textwidth]{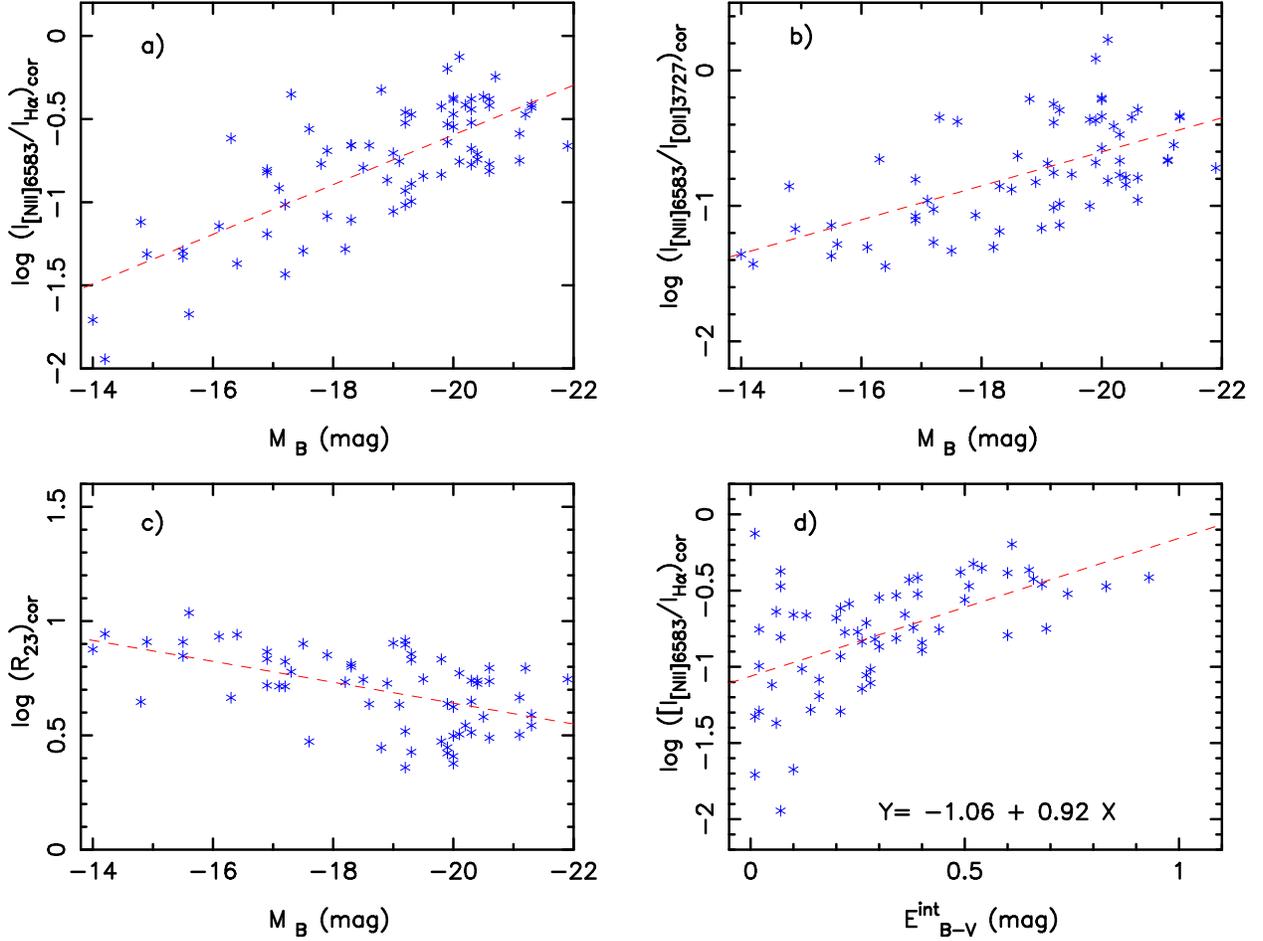}
\caption{
Intrinsic emission line flux (I$_{\lambda}$, reddening and underlying 
absorption were corrected)
ratios as a function of the galaxy absolute blue magnitude $M_B$ 
and the color excesses, $E^{int}_{B-V}$. a) \nii6583/\ha;  b) 
\nii6583/\oii3727; c) $R_{23}$ = (\oii3727+\oiii4959, 5007)/\hb; 
d) \nii6583/\ha\ as a function of $E^{int}_{B-V}$.}
\label{fig-metal}
\end{figure*}

Numerous studies have claimed the existence of a 
metallicity-luminosity relation in a variety of classes of galaxies: 
dynamically hot galaxies, i.e. ellipticals, bulges, and dwarf 
spheroidals, dwarf \hii\ galaxies, irregular galaxies and spirals 
(Lequeux et al. 1979; Skillman, Kennicutt, \& Hodge 1989; 
Stasi{\' n}ska \& Sodr{\' e} 2001).  In Figure 
\ref{fig-metal}a--\ref{fig-metal}c, we show the various 
metallicity indices, $R_{23}$=(\oii3727+\oiii4959, 5007)/\hb\ 
(Pagel et al. 1979), \nii6583/\ha\ (van Zee et al. 1998) and 
\nii6583/\oii3727 (Dopita et al. 2000) as a function of the total 
absolute blue magnitude $M_B$.  The line flux ratios were corrected 
for both internal (using the value of $E^{int}_{B-V}$) and Galactic 
extinction, and underlying stellar absorption.  Since the 
reddening correction becomes more uncertain for galaxies with small 
$EW$(\ha) and $EW$(\hb), we only use those objects that have $EW$(\hb) 
$>$ 5 \AA\ and thus the most reliable reddening corrections.  We 
do find a good correlation between $M_B$ and the metallicity indices 
\nii6583/\ha\ and $R_{23}$. The correlation of the 
\nii6583/\oii3727 index with $M_B$ is statistically significant, 
but at a rather low level. These metallicity indices show clear 
trends with galaxy absolute magnitude, confirming that indeed there 
is a relation in blue compact galaxies between the overall metallicity 
of the star forming regions and the galaxy luminosity. The higher the 
galaxy absolute magnitude, the higher the heavy element content.  
This relation suggests that the metallicity of faint, low mass BCGs 
is low.

We now examine whether there is a relation between the color excesses 
due to internal extinction, $E^{int}_{B-V}$, and the overall 
metallicity of the galaxies. So far, there have been contradictory 
claims in this respect.  Zaritsky et al. (1994) found no evidence for 
a systematic dependence between reddening and abundance in a sample of 
39 disk galaxies.  In other contexts, Stasi{\' n}ska \& Sodr{\' e}(2001) 
found that the nebular extinction as derived from the Balmer decrement 
strongly correlates with the effective metallicity of the emission line 
regions of spiral galaxies.

Figure \ref{fig-metal}d shows $E^{int}_{B-V}$ as a function of the 
metallicity indicator \nii6583/\ha, which is less affected by the 
reddening correction. We find there is a clear correlation, 
${\rm log}(I_{{\rm [NII]}6583}/I_{{\rm H}\alpha})_{cor} = - (1.06 
\pm 0.07) + (0.92\pm 0.17) E^{int}_{B-V}$. \nii6583/\ha\, tends to be 
larger 
for larger values of $E^{int}_{B-V}$, when $E^{int}_{B-V} > 0.1$. 
Internal extinction indeed correlates with the overall metallicity 
of BCGs, especially among the galaxies with large $E^{int}_{B-V}$. 
Since the metallicity indices correlate with $M_B$, the correction
between $I_{{\rm [NII]}6583}/I_{{\rm H}\alpha}$ and  $E^{int}_{B-V}$ 
also suggest
the internal extinction of brighter, more massive BCGs is higher.

\section{Summary}

In this paper, we have measured the fluxes and equivalent widths 
of emission lines, the fluxes at several points of the continuum, 
the 4000 \AA\ Balmer break index, and the equivalent widths of 
absorption lines for our BCGs sample.  Then we have analyzed the 
fluxes and equivalent widths of emission lines for the star-forming 
galaxy subsample.  

Our main results are:
\begin{enumerate}
\item
We have measured the fluxes and equivalent widths of emission lines 
for BCGs, and estimated their errors. The typical uncertainties 
of the measurements are less than 10\% for the brightest \ha6563, 
\hb4861, \oii3727, \oiii4959,5007, \nii6583, \sii6717,6731 lines 
and less than 20\% for the weaker \hg4340, \hei5876, \oi6300\, and 
\oiii4363 lines.  Our line intensity ratios are in good agreement with 
those of previous studies. The color excesses due to internal 
extinction was calculated based on the flux ratio of \ha/\hb.

\item
The equivalent widths of absorption features, the continuum colors 
and the 4000 \AA\ Balmer break indices of BCGs were measured. The mean 
certainty in these measurements is better than 15\%.

\item
The equivalent width of \ha\, emission line is correlated with $M_B$; 
lower luminosity systems tend to have larger equivalent widths. This 
can be 
explained by assuming blue galaxies dominate the faint SFG 
population, while redder galaxies dominate the bright SFG 
population.

\item
On average, luminous BCGs are likely to be enhanced in nitrogen 
abundance. This suggests that in faint, low-mass, BCGs nitrogen is 
a primary element, whereas in brighter, more massive BCGs nitrogen 
comes from a secondary source.

\item
A correlation is found between most of emission lines, the strongest 
correlations are found between \oii3727, \hg, \hb\ and \ha.  
Besides \oii3727, \hg\ and \hb\ lines can serve as a reliable star 
formation tracers for strong emission line galaxies.

\item
Metallicity indices show clear trends with galaxy absolute 
magnitude, confirming that there is a relation in galaxies between 
the overall metallicity of the star forming regions and the galaxy 
luminosity. Faint, low-mass BCGs have lower metallicity and internal 
extinction.

\end{enumerate}

\begin{acknowledgements}
We thank the referee Dr. Y. I. Izotov for his valuable comments, 
constructive suggestions on the manuscript.  This work is based on 
observations made with the 2.16m telescope of the Beijing 
Astronomical Observatory(BAO) and supported by the Chinese National 
Natural Science Foundation (CNNSF 10073009).   
S.C. thanks the Alexander von Humboldt Foundation, the Federal 
Ministry of Education and Research, and the Programme for Investment in 
the Future (ZIP) of the German Government for support.
X.K. has been financed by the Special Funds for Major State Basic 
Research Projects of China and the Alexander von Humboldt Foundation 
of Germany.
\end{acknowledgements}

Tables 1, 2, 3, 4, 6 are only available in electronic form at the
CDS via anonymous ftp to cdsarc.u-strasbg.fr (130.79.128.5) or via
http://cdsweb.u-strasbg.fr/cgi-bin/qcat?J/A+A/

\end{document}